\documentclass[letterpaper, 10 pt, conference]{ieeeconf}  

\IEEEoverridecommandlockouts                              

\title{\LARGE \bf
\textcolor{black}{Preliminary Experimental Results of Context-Aware Teams of Multiple Autonomous Agents Operating under Constrained Communications}
}

\author{\textcolor{black}{Jose Martinez-Lorenzo, Jeff Hudack, Yutao Jing, Michael Shaham, Zixuan Liang, } \\ \textcolor{black}{Abdullah Al Bashit, Yushu Wu, Weite Zhang, Matthew Skopin, Juan Heredia-Juesas,  }\\ \textcolor{black}{Yuntao Ma, Tristan Sweeney, Nicolas Ares, Ari Fox}
\thanks{\textcolor{black}{College of engineering, Northeastern University, Boston, MA 02115, USA, j.martinez-lorenzo@northeastern.edu }
}
}

\usepackage{makecell}
\usepackage{verbatim}
\usepackage{graphics} 
\usepackage{epsfig}
\usepackage{color}
\usepackage{amssymb}

\usepackage{array,amsmath,booktabs}
\usepackage{float}
\usepackage{cite}
\usepackage{lipsum}
\usepackage{algorithm}

\usepackage{dblfloatfix}

\usepackage[noend]{algpseudocode}

\usepackage{multicol}
\usepackage{amsmath}

\usepackage{todonotes}

\usepackage{tikz}
\usepackage{mathtools}
\usepackage{array}

\DeclareRobustCommand\sampleline[1]{%
    \tikz\draw[#1, thick] (0,0) (0,\the\dimexpr\fontdimen22\textfont4\relax)
    -- (1em,\the\dimexpr\fontdimen22\textfont4\relax);%
}

\newcolumntype{M}[1]{>{\centering\arraybackslash}m{#1}}

\definecolor{mygreen}{rgb}{0.0,0.4,0.0} 
\definecolor{mycyan}{rgb}{0.0,0.72,0.92} 
\definecolor{myyellow}{rgb}{0.71,0.65,0.26}
\definecolor{mymagenta}{rgb}{0.54,0.17,0.89}

\begin{document}

\maketitle
\thispagestyle{empty}
\pagestyle{empty}

\begin{abstract}
This work presents and experimentally test the framework used by our context-aware, distributed team of small Unmanned Aerial Systems (SUAS) capable of operating in real-time, in an autonomous fashion, and under constrained communications. Our framework relies on three layered approach: (1) Operational layer, where fast temporal and narrow spatial decisions are made; (2) Tactical Layer, where temporal and spatial decisions are made for a team of agents; and (3) Strategical Layer, where slow temporal and wide spatial decisions are made for the team of agents. These three layers are coordinated by an ad-hoc, software-defined communications network, which ensures sparse, but timely delivery of messages amongst groups and teams of agents at each layer even under constrained communications. Experimental results are presented for a team of 10 small unmanned aerial systems tasked with searching and monitoring a person in an open area. At the operational layer, our use case presents an agent autonomously performing searching, detection, localization, classification, identification, tracking, and following of the person, while avoiding malicious collisions. At the tactical layer, our experimental use case presents the cooperative interaction of a group of multiple agents that enable the monitoring of the targeted person over a wider spatial and temporal regions. At the strategic layer, our use case involves the detection of complex behaviours--i.e. the person being followed enters a car and runs away, or the person being followed exits the car and runs away--that requires strategic responses to successfully accomplish the mission.      
\end{abstract}

\section{Introduction}

Recent advancements in the fields of Artificial Intelligence \cite{Aydemir:A2, Aydemir:A, FRcnn, dosovitskiy2015flownet, fu2016one, lillicrap2015continuous, zhu2018reinforcement}, Machine Learning \cite{lecun2015deep, ngiam2011multimodal, yang2017deep, srivastava2012multimodal}, Robotics \cite{whitney1987historical, peng2018sim, Diankov:R, Sjoo:K, de2018integrating, cifuentes2016probabilistic}, and Signal Processing \cite{WeiteAPS2019, tirado2018towards, zhang2018single} have provided humankind with unique set of tools that, for the first time in history, have the potential to address some of the most important problems existing in the field of group autonomy of unmanned systems \cite{Koenig:N}. Nowadays, group autonomous systems require either direct human control of many systems \cite{finn2017deep, sinapov2014learning}, contract and auction techniques \cite{kensler2009neural, liekna2012experimental}, and or coalition methods \cite{shehory1995task, li2004stable, shehory1996formation}. The latter are heavily dependent on the communications channel, which is often constrained in many realistic scenarios. Other approaches based on Markov Decision Processes do not scale linearly with the number of agents and states, and they often result in a slow reaction to unexpected events, \cite{Kaelbling:L, katt2017learning, ross2011bayesian, Li:J, Littman95, Ross:S, NIPS2010_4031, Silver:2010:MPL:2997046.2997137}.

\begin{figure*}[]
	\centering
	\includegraphics[scale=.53, trim = 0mm 50mm 0mm 0mm, clip]{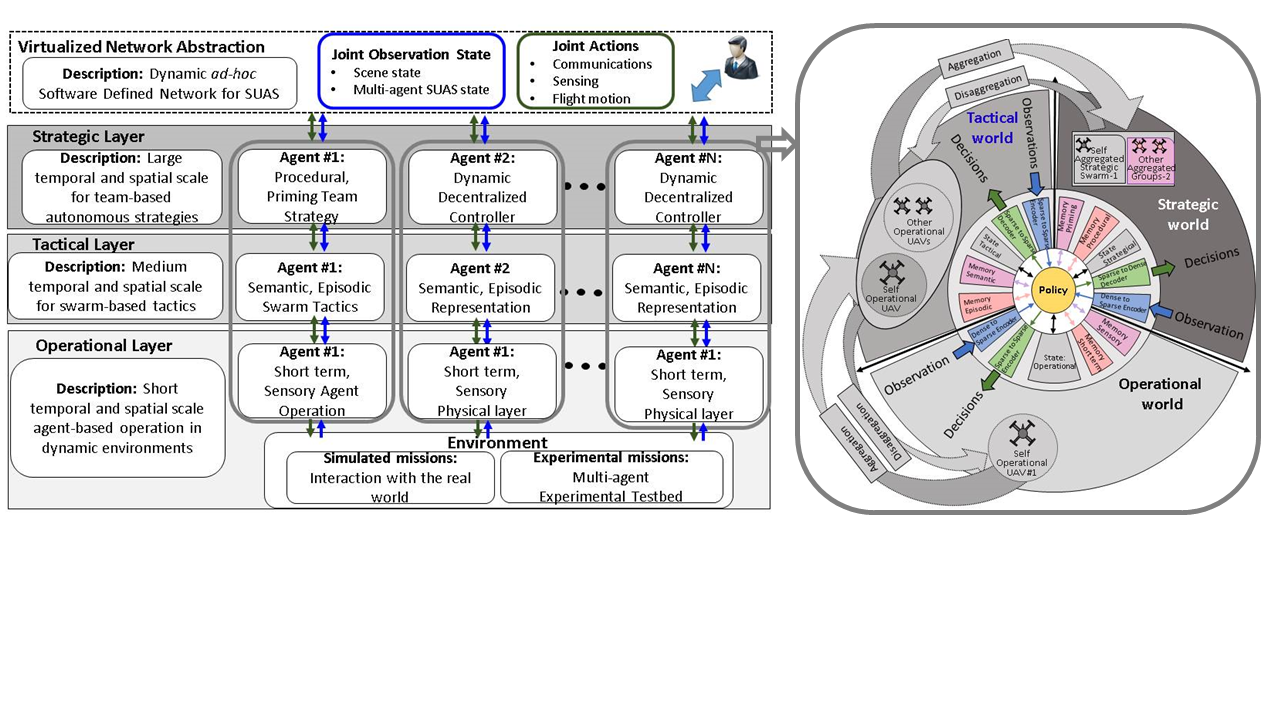}
	\caption{Decentralized autonomous systems: (left) team operator perspective; (right) agent in the team perspective}
	\label{system_architecture}
\end{figure*}

This paper describes and experimentally validates the framework---hardware, software, and system of systems architecture---used by our context-aware, distributed team of Small Unmanned Aerial Systems (SUAS) to be able to operate in real-time, in an autonomous fashion, and under constrained communications. Our framework relies on three layered approach: (1) Operational layer (fast temporal and narrow spatial scale; partially mimicking functionality of human's peripheral nervous system) - here a single agent performs on-board detection, localization, classification, identification, tracking, following while avoiding malicious collisions; this layer relies on hardware and software that enable to fuse and sparsify in real-time 4D full motion video, 4D millimeter wave radars, 4D infrared cameras using Deep Learning and 4D (space + time) Compressive Sensing (CS); (2) Tactical Layer (intermediate temporal and spatial scale; partially mimicking functionality of human's muscular system): here a group multiple autonomous agents collaborate to jointly perform a complex task that cannot be executed by a single agent due to their spatial (navigation) and temporal (perception) limitations; and (3) Strategical Layer (slow temporal and wide spatial scale; partially mimicking functionality of the endocrine system): here teams of multiple autonomous agents cooperate to jointly perform a multi-step complex task that cannot be executed by a group of autonomous agents due to their spatial (navigation), temporal (perception), and energy (endurance) limitations. These three layers are coordinated by an {\it ad-hoc}, software-defined communications network, which ensures sparse, but timely delivery of messages amongst groups and teams of agent even under constrained communications.

\section{System Architecture}
The hierarchical architecture adopted by our autonomous multi-agent system is the one presented in Fig. \ref{system_architecture}. From an operator perspective, see the left side of the figure, a general mission is specified for a team of agents that must organize themselves to accomplish a particular mission. In this paper use-case, the general mission is defined as {\it search and monitor people in a given region}. Based on this mission, an off-line planner parses a multi-layer policy (controller) to each agent in the network using a top-to-bottom approach. The latter leverages on the use of a set of {\it memory banks}, which resemble the different types of memories used by the human body, including: {\it (i)} long term strategic, which covers spatial priming memory and temporal procedural memory; {\it(ii)} long term tactical, which covers spatial semantic memory and temporal episodic memory; and {\it(iii)} short term memory and  sensory memory. The strategic memory is used to load the initial strategic policy, as well as the type of decisions and observations available for the team of agents at this level. A similar functionality is provided to the tactical and operational memory, regarding decisions, observations  and policies at its corresponding layer. From an agent perspective, our architecture enables the each unmanned system to reason about its own operation, its tactical relationships with a subgroup of agents with whom it is cooperating in a joint task, and its strategical contribution to the overall mission. This perspective is shown on the right part of Fig. \ref{system_architecture}, and a thorough description is described in the next section.

\begin{figure*}[htp]
	\centering
	\includegraphics[scale=.7, trim = 0mm 20mm 0mm 0mm, clip]{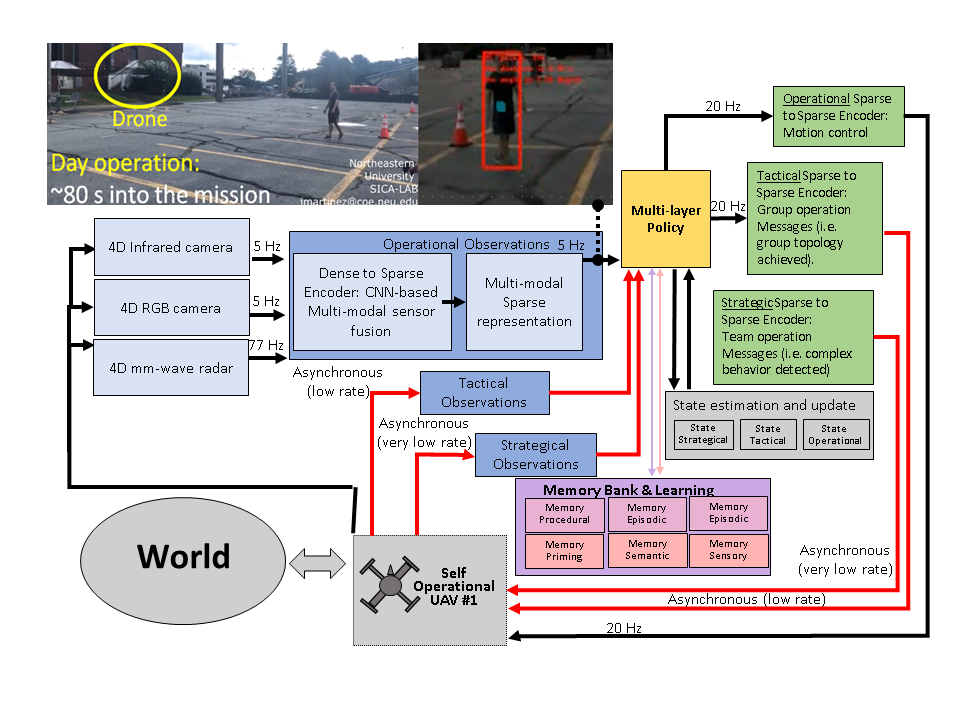}
	\caption{Architecture of multiple sensors fusion with Convolutional Neural Network. Inference results of the fusion module feed into our drone policy,  which controls the motion of the drone or performs complex behaviors and output environmental prediction.}
	\label{multimodal_representation}
\end{figure*}

\begin{figure}[htp]
	\centering
	\includegraphics[scale=.5, trim = 0mm 50mm 10mm 0mm, clip]{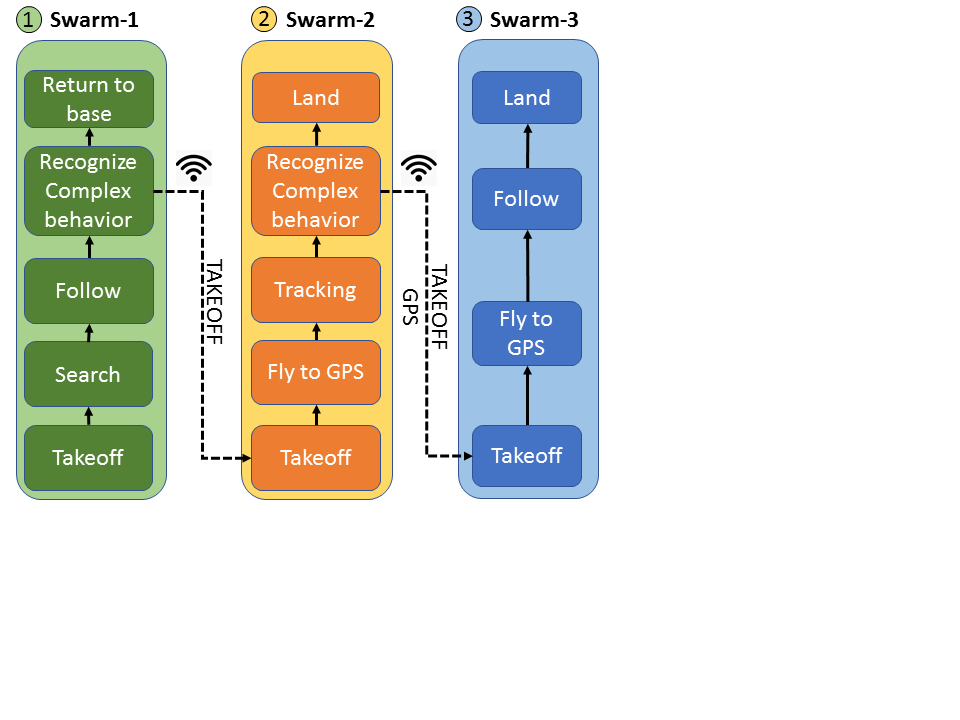}
	\caption{Multi-Swarm mission description}
	\label{tasks}
\end{figure}




\begin{figure*}[htp]
	\centering
	\includegraphics[scale=.54, trim = 0mm 60mm 0mm 0mm, clip]{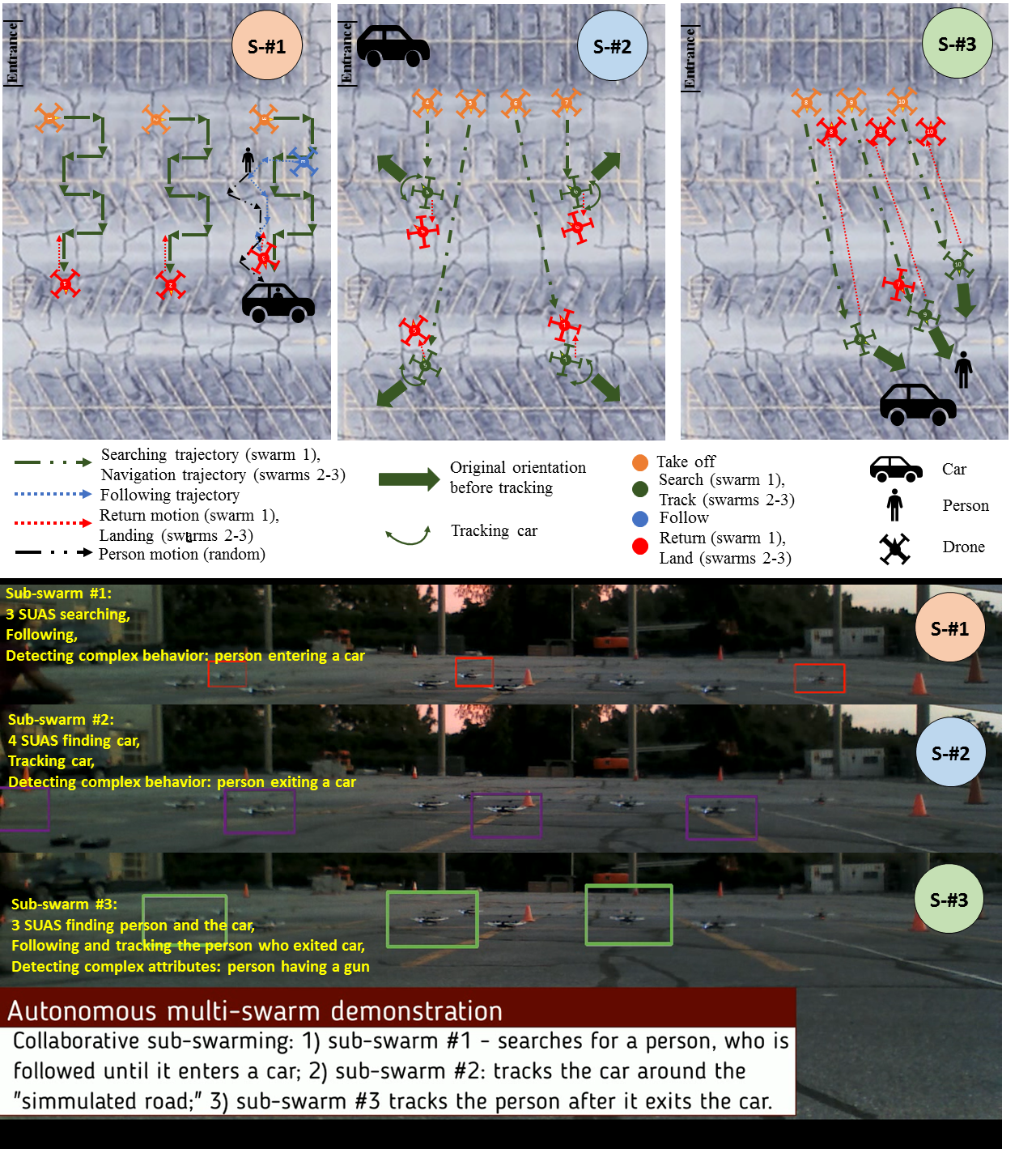}
	\caption{Autonomous Multi-swarm demonstration. 1) Sub-swarm $\#1$ searches for a person, who is followed until it enters a car; 2) Sub-swarm $\#2$ tracks the car around the "simulated road"; 3) Sub-swarm $\#3$ tracks the person after it exits the car.}
	\label{All_swarms}
\end{figure*}

\begin{figure*}[htp]
	\centering
	\includegraphics[scale=.69, trim = 0mm 20mm 0mm 20mm, clip]{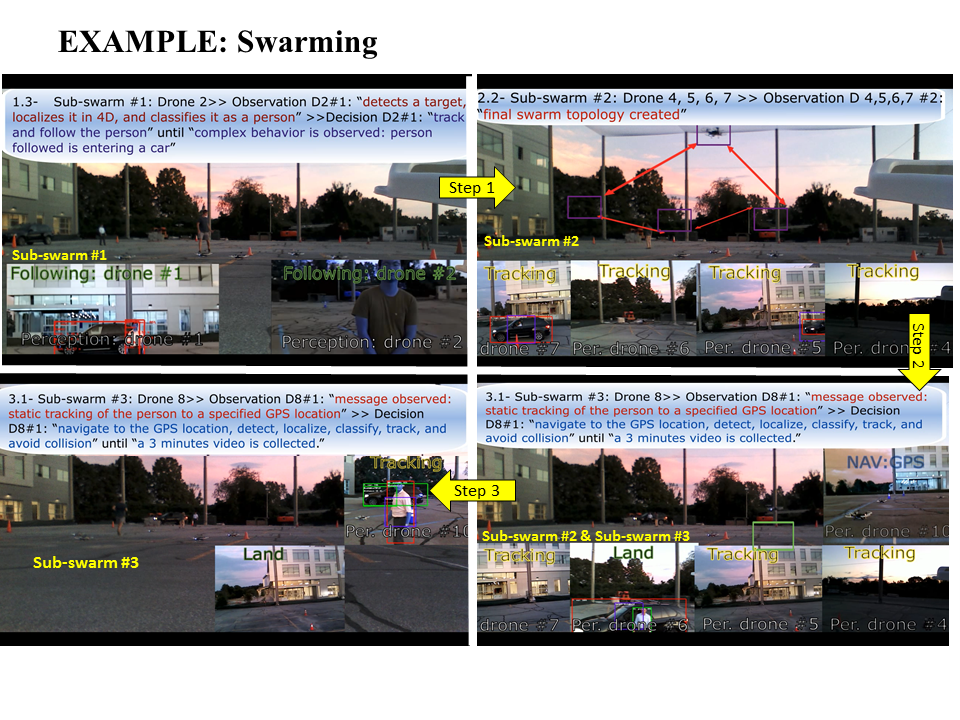}
	\caption{Overall sequence of the whole mission recorded by a stationary ground camera and a drone camera. }
	\label{whole_mission}
\end{figure*}


\section{Multi-layer perception}
As shown in Fig. \ref{multimodal_representation}, the agents (SUAVs) in our network can be equipped with  three different type of perception sensors: 4D RGB camera, 4D Infrared camera, and a 4D mmWave radar. At the operational level, the raw data of the three sensors is parsed into a Vector Processing Unit, which runs a fine-tuned Convolutional Neural Network to perform the sensor fusion and to provide a sparse representation of the scene. As it can be seen on the top area of Fig. \ref{multimodal_representation}, our dense to sparse perception module is capable of outputting sparse information about the scene at a 5 Hz rate-- an enhanced frame rate of 100 to 1000 Hz should be achieved with our current architecture. This output contains a list of targets in the scene (e.g., person, car, etc.), classification confidence level for each target, targets' bounding boxes in 2D,  targets' ranges from the agent, targets' angular location relative to the agent's orientation, as well as 4D GPS Geo-location of both targets and agent. At the tactical level, medium priority observations involving other agents within the same group, jointly performing a particular activity, is sparsely parsed through the {\it ad-hoc} network in an asynchronous fashion at a reduced average rate ($\sim$ 0.01 Hz per mission). Similarly, at the strategical level, top priority observations of events that require an update on the team strategy are parsed through the {\it ad-hoc} network in an asynchronous fashion at very low average rate ($\sim$ 0.005 Hz per mission).   

\section{Multi-layer policy}

The vector targets at the operational layer is synchronously parsed to our multi-layer policy block (see Fig. \ref{multimodal_representation}), which uses a sparse to sparse motion controller to generate motion trajectories based on the agent's particular state. At the operational level, each drone can simultaneously be active in one or more of the following operational states: {\it idle/sleeping}, {\it takeoff/landing}, {\it searching}, {\it following}, {\it tacking}, {\it Navigating to a GPS location}, {\it returning to base}. The latter sparsification at the operational state affords scalability of the Multi-layer policy. At the tactical level, the multi-layer policy handles the information received either by its own operational observations or from another member of its tactical group. The policy enables an asynchronous coordination of the group of agents at the tactical level. When a group of agents are not able to continue a particular group activity, the strategic policy may be able to recruit another group of agents that can finalize the mission in a suitable fashion. The strategic policy observes and controls the strategic perception and actuation channels. The use-case described below will clearly emphasize the type of observations and actions that are provided for each one of the components of the multi-layer policy.

\section{Multi-layer decisions}
The multi-layer policy creates a sparse vector that encodes the actions needed at the strategical, tactical, and operational level. In the latter, the {\it Operational sparse to sparse encoder} shown in Fig. \ref{multimodal_representation} generates the signals needed to control the lower-level motion controller. Our system follows a control approach similar to the one presented in \cite{lee2018making}. Specifically, once target is recognized and localized, the drone will change its state to follow or track the object and update its position, $\textbf{p}_t$, based on the change in position, $\Delta\textbf{p}$, obtained from its own observations. Position updates can be made by sending the flight controller either local velocity setpoints or local or global position setpoints. By receiving the angle and the distance of the object relative to its current position and orientation, the controller will decide how much to rotate and how to adjust its position. Various uncertainties, $\mathcal{U}$, like external forces, ${J}$ (e.g., wind), can affect the motion of the drone, which the flight controller needs to be capable of compensating for. Changes to the controller can also come from communication with other drones or other swarms, or from recognizing complex behavior or patterns. When a drone recognizes certain behavior occurring among the objects it is seeing (e.g., a person entering a car), it can communicate to the other drones to change their state (e.g., to return home) and to other swarms to begin or change their mission.

\begin{figure}[htp]
	\centering
	\includegraphics[scale=.42, trim = 1mm 50mm 40mm 0mm, clip]{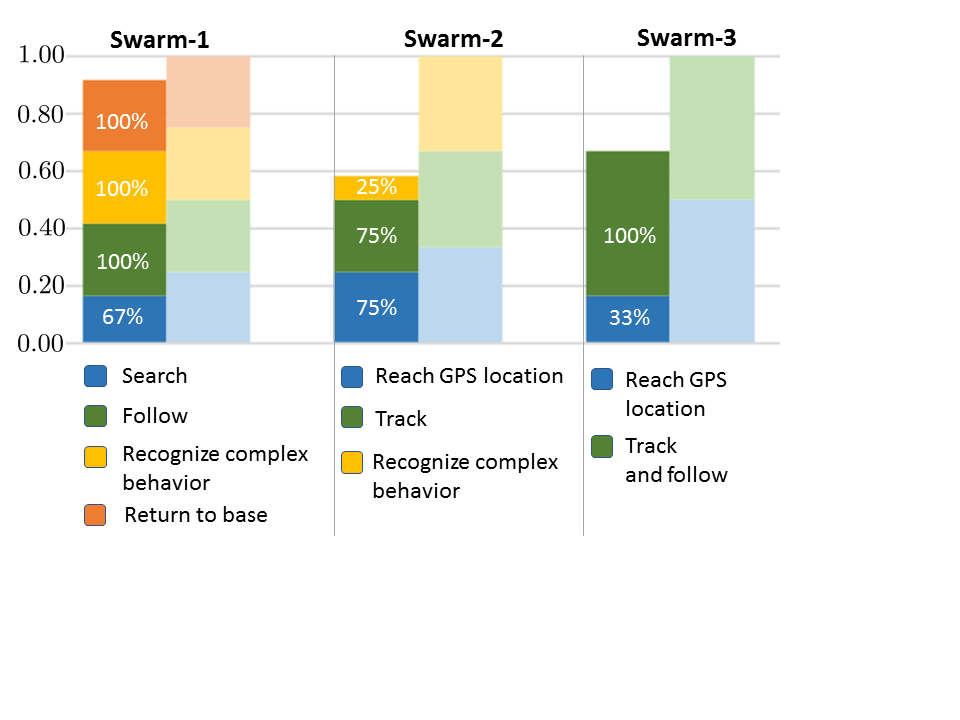}
	\caption{Multi-swarm performance. For each swarm, the right bar is desired performance for each task, while the left bar is the actual performance.}
	\label{success}
\end{figure}

\section{Results}

At the top of the system is the mission controller, which controls the subsystems in an attempt to achieve the swarm's objective. Mission objectives for the swarm of drones is typically defined by an area of exploration and a searching objective, e.g., find survivors in a disaster-struck area. To maximize the ability to search an area and understand the environment, the swarm needs to be divided into a specific number of subswarms depending on the environment and objective. For example, the area of exploration can be partitioned into different sections, each of which is searched by a different subswarm. If needed, subswarms can decide to split up into smaller subswarms depending on what is best for the environment it is in. For example, a subswarm may encounter a building or multiple buildings, and need to split up to search these newly encountered parts of the environment. On the smallest scale in this system, individual drones make observations and act on them based on a learned policy. Communication with other drones in the subswarm occurs depending on its observations. A mission can be ended when the mission level system sends a signal, either based on time or observations, that the mission is over. 

This paper brings up the proof of concept of experimenting with drones in navigating, tracking, following, and landing modes with a swarm of ten drones, as represented in Fig. \ref{tasks}. In this experiment, swarm-one, swarm-two and swarm-three have three, four, and three drones, respectively, who are participating in the mission, as it is represented in detail in Fig. \ref{All_swarms}. The tasks for swarm-one are detecting, following the person until he/she is entering into the car (which has been defined as a complex behavior), and finally return to base. The tasks for swarm-two are flying to the predefined GPS locations, and start tracking, which means facing to a direction where the car moves. Until the car stopped and the person went out of the car, which also has been defined as a complex behavior, or received the LAND command from both peers or ground control station, the drones keep on tracking. In this stage, if any of the drone detects the complex behavior, i.e. a person and a car is present, it immediately sends ARM command to the swarm-three to fly around the sending drone. The tasks for swarm-three are flying to the GPS position sent from the drone, and start following the person. The swarm-three looks the person and maintain a constant distance with the person before coming back to the base station. The detailed sequence of the mission, along with some frames captured by the cameras of the drones incorporating their perception, is shown in Fig. \ref{whole_mission}.

The perception in swarms one and two, which leads to the detection and tracking of the person and car, is performed by a computer vision algorithm based on the pre-trained MobileNet-SSD convolutional neural network \cite{howard2017mobilenets}. The front RGB camera captures the target within a rectangle. When the midpoint of that rectangle appears deviated from the center of the camera, the flying algorithm promotes the drone to rotate until the target center point meets within the threshold of the tracking pattern. Meanwhile, the depth camera measures the average distance to the target. When the distance increases over a given value $D_0$, the flying algorithm pushes the drone closer to the target and vice-versa. 

On the other hand, the perception in swarm three, which leads to the following at a constant distance of the person, is performed by extracting the range distance from the radar 3D point-cloud followed by a negative feedback to the UAV fight controller and moving $-(R-R_0)$ meters in the range direction (the direction of the front-view of the camera), where $R_0$ is the constant following distance and $R$ is the detected range distance by the radar.

Figure \ref{success} shows the performance of the current experiment. For swarm-one, $67\%$ of drones (two out of three) found and detected the person successfully, made the decision to follow them, and finished the following task successfully; it neither lost the target during following nor hit the obstacle accidentally. The other drone failed to finishing its tasks successfully without finding the person; However, at the end, all the drones in swarm-one received the LAND command and returned to base successfully. In the swarm-two, $75\%$ of drones (three out of four drones) reached their predefined GPS points, rotated around their own z-axis and tracked the car as excepted, and one of these three drones finished the complex task, which was defined as detecting the target person entering into the car, and then sending messages to swarm-three to arm and take off. Finally, in the swarm-three, $33\%$ of drones (one out of three drones) received the TAKEOFF message from swarm-two successfully, and done the mission to fly over the GPS point, tracking as well as following tasks successfully. However, the other two did not take off as being supposed to.

\subsection{Discussion}
While testing, if the drone keeps navigating, tracking, following and landing, then the tasks are considered as successful, and they are defined as performed the expected mission. It is observed that tracking in negative areas---such as the dark side of the car---, communication antenna orientation, wind speed, sensor calibration, distance between drones resulting packet loss affect the detection, navigating, and tracking performance for the swarms to perform a desired task.

In addition, when multiple targets are captured by the camera of the drone, such as several people or cars in the same frame, some constraints may limit the drone operation, leading to a possible false tracking. In the presented case, the person or car that first appears in the drone's field of view is considered as the main target, tracking it without losing or switching it. However, if two people appear on the scene too close or lap over each other, it is possible that the tracker switches the main target, leading to a failed mission. In future experiments, where the requested mission will be much more complex than current experiment, 
it will be crucial for the team of SUAS to obtain as much as information as possible from the outside environment. For these cases, a multiple objects tracking (MOT) approach is expected to be more reliable in realistic scenarios. This functionality, which will be vital for a team of SUAS to perceive a large-scale environment, is already available with current online MOT methods, such as deepSORT, MHT\_bLSTM, and OneShotDA, benefiting the extensibility of our approach to more complex missions. 

Moreover, the current mm-wave radar is employed using time-division multiplexing where only one Tx is transmitting at a time, resulting in a possible low signal-to-noise ratio (SNR) at the receiver end and causing a poor detection accuracy if the object is too far away. Future radar architecture will use spatial multiplexing schemes such as binary-phase-modulation to perform the detection, where all the Txs are transmitting simultaneously to achieve a much higher SNR at the receiver end.

\section {Conclusion}
This paper has shown an experimental test of a context-aware distributed team of SUAS coordinately working on a multi-step complex mission, capable of operating in real-time, in an autonomous fashion, and under constrained communications. In this experiment, 10 drones divided into three teams perform the complex three-step task of (i) searching, detecting and following a person until enters into a car, (ii) navigating to a specific GPS position and tracking a car until a person leaves the car, and (iii) navigating to a GPS position given by the previous team, and follow a person at a constant distance for a period of time. The proposed framework relies on a three layers approach: operational, tactical, and strategical, corresponding to single agent actions, group of agents collaboration, and teams of multiple groups of agents join cooperation, respectively. The complex mission is carried out based on the continuous loop perception--policy--decision architecture. The perception is done based on the fusion of 4D RGB and 4D infrared cameras, together with 4D mmWave radar, and the communication among the agents in the teams is performed by and ad-hoc network. 
The experimental validation showed that the complex task was propitiously achieved by the cooperation of the three teams. Although some agents in the teams may have not had the expected behaviour due to possible packages loss, non-optimal illumination conditions for detection and tracking, and navigation issues due to the uncertainties, the global behaviour of the swarm managed to successfully complete the required mission.


\vspace{1mm}

{\bf \noindent Acknowledgements.} This work is funded by the U.S. Air Force Research
Laboratory (AFRL), BAA Number: FA8750-18-S-7007.

\bibliographystyle{IEEEtran}
\bibliography{ICRA19_Amato_Refs,References_ICRA2020}

\end{document}